\documentclass[a4paper]{article}

\usepackage{icrc2013,xspace}
\usepackage{color}
\title{A Central Laser Facility for the Cherenkov Telescope Array}

\shorttitle{A CLF for the CTA}

\authors{
M. Gaug$^{1,2}$,
C. Aramo$^{3}$,
M. Cilmo$^{3,4}$,
F. Di Pierro$^{5}$,
A. Tonachini$^{6}$,
P. Vallania$^{5}$
for the CTA Consortium.
}

\afiliations{
$^1$ F{\'i}sica de les Radiacions, Departament de F{\'i}sica, Universitat Aut{\`o}noma de Barcelona, 08193 Bellaterra, Spain. \\
$^2$ CERES, Universitat Aut{\`o}noma de Barcelona-IEEC, 08193 Bellaterra, Spain. \\
$^3$ INFN, Sezione di Napoli, Via Cintia 2, 80126 Napoli, Italy. \\
$^4$ Dipartimento di Scienze Fisiche, Universit\`a degli Studi di Napoli Federico II, Via Cintia 2, 80126 Napoli, Italy.\\
$^5$ Osservatorio Astrofisico di Torino
     dell'Istituto Nazionale di Astrofisica \& INFN-Torino, via P. Giuria 1, 10125 Torino, Italy. \\
$^6$ Dipartimento di Fisica dell'Universit\`a di Torino \& INFN-Torino, 
     via P. Giuria 1, 10125 Torino, Italy. \\
}

\email{markus.gaug@uab.cat}

\abstract{A Central Laser Facility is a system often used in astroparticle experiments based on
arrays of fluorescence or Cherenkov light detectors. The instrument is based on
a laser source positioned at a
certain distance from the array, emitting fast light pulses in the vertical
direction with the aim of calibrating the array and/or measuring the atmospheric transmission.
In view of the future Cherenkov Telescope Array (CTA), a similar device
could provide a calibration of the whole installation, both
relative, i.e. each individual telescope with respect to the rest of the array, and
absolute, with a precision better than 10\%, if certain design
requirements are met. Additionally, a precise monitoring of the
sensitivity of each telescope can be made on time-scales of days to years.
During calibration runs of the central laser facility, all detectors will be
pointed towards the same portion of the laser beam at a given altitude.
Simulations of the possible configurations of a Central Laser Facility for
CTA (varying laser energy, pointing height and distance from the
telescopes)  have been performed.}

\keywords{Central Laser Facility, CTA, Instrumentation and Methods for Astrophysics, IACT}

\begin{document}
\maketitle

\section{Introduction}

Central Laser Facilities (CLF) have been used widely to calibrate fluorescence detectors~\cite{HiRes2006,auger2006,auger2011,auger2013,ta2009,ta2012}. 
Such facilities as this employ a laser to emit fast light pulses of precisely monitored power, mostly in vertical direction. If properly depolarized, the scattered laser light 
received by the focal plane detectors resembles that from fluorescing ultra high energy cosmic ray shower tracks and is
used to calibrate the response of the photo-detectors to these. 

In the case of Imaging Atmospheric Cherenkov Telescopes (IACTs)~\cite{weekes2005,buckley2008,hinton2009,holder2012,cta}, 
a part of the laser path is seen as a track traveling across the focal plane of the camera, on time scales of micro-seconds. 
IACTs
 are optimized for the observation of Cherenkov light from air showers that are observed head-on
and yield light pulses with full-width-half-maxima of typically few nanoseconds. 
Telescope calibration by a CLF is then only useful if 
the light pulses from CLF tracks are amplified and electronically transmitted and digitized in the same, undistorted way as the 
shorter Cherenkov light pulses. 
In this case, a CLF can be used to monitor the sensitivity of each individual telescope, including mirrors and camera, 
and to cross-calibrate telescopes, or telescope types, between each other.
Contrary to the already existing CLFs, we propose to operate a CLF at different wavelengths, allowing for a full spectral characterization 
of each telescope. 
Such a calibration scheme has the advantage to be fast and relatively cheap, as only one device is involved for the entire array. 

VERITAS is the only IACT that has explored calibration of its Cherenkov telescopes with the help of a CLF~\cite{veritas2005,veritas2008}. 
All installations make use, however, of other calibration devices and rely on their CLFs to yield redundant information. 
The Auger experiment uses, moreover, two laser facilities and four LIDAR stations to characterize the atmosphere and the fluorescence detectors~\cite{auger2013}. 

Given the experience of these installations, a possible use of a CLF for the future Cherenkov Telescope Array (CTA)~\cite{cta,ctaconcept,ctamc} is discussed here. 
The CTA will consist of a Southern installation, covering about 10~km$^2$ with telescopes, and a Northern one of about 1~km$^2$ extension. 
While the Southern array will contain at least three 
different telescope types, employing different mirror sizes and fields-of-view, the smaller Northern array will consist of two types of telescopes. 
In both cases, large-size telescopes will be located in the center, surrounded by medium-size and small-size telescopes, the latter only for the Southern array.

\section{Geometrical considerations}

The basic idea of a calibrating device for a telescope array is that all telescopes observe the same light source, in our case the same part of a laser beam. 
With an array of telescopes distributed over an area of 1--10~km$^2$, this is obviously impossible, however one can try to make them observe a 
very similar part of the laser beam, and to apply only small corrections for each telescope. 

Each telescope camera observes the scattered light from the laser beam, within a path length defined 
by the field-of-view (FOV) of the camera. The beam is then seen in the camera traveling as a stripe from the uppermost part of the camera down to the lowest part.
Each pixel along that line will observe a part of the beam corresponding to its FOV. 
The accumulated charge in an illuminated pixel reflects then the output power of the laser, scattering and absorption of laser light in 
the atmosphere and the telescope sensitivity to light at the laser wavelength. If the first two parts can be controlled to a precision better 
than the known telescope sensitivities,
the CLF can ultimately serve for an absolute calibration of each telescope or the whole observatory. 



The geometry of the setup has to be chosen such that each camera 
observes roughly the same part of the laser beam.
At the same time, one should avoid to observe the beam at a place where scattering of light is strongly influenced by aerosols, 
i.e. the observed part of the laser beam has to be always above the Planetary Boundary Layer, or above the Nocturnal Boundary Layer, if present. 
Hence a minimum height of $\sim$2~km is required, depending on the atmospheric conditions on site. 
One could now think of a very close device, even one situated at the very center of the CTA observatory. However, in this case the closest 
telescopes will observe an almost infinite part of the laser beam, since the cotangent of the zenith angle of observation and the FOV of 
each pixel are involved. On the other hand, a very distant device, observed at a high zenith angle, makes the observed laser path 
long and extending to very high altitudes. In the end, an optimum distance has to be found somewhere in between. 
\par\noindent
\begin{figure}[h!]
\centering
\includegraphics[width=0.49\textwidth]{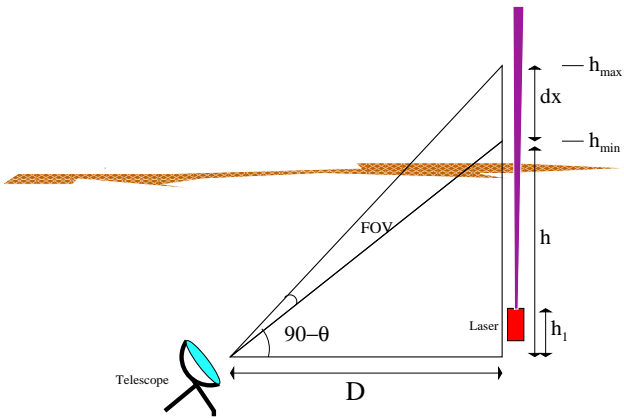}
\caption{Sketch of the introduced geometry.
The brown band in the 
background sketches the nocturnal boundary layer.\label{fig1}}
\end{figure}

Figure~\ref{fig1} shows a sketch of the introduced geometry:  a pixel of one CTA camera, or a complete camera, sees the laser beam above a height $h$ from ground, 
under a zenith angle $\theta$. 
If the telescope is located at a distance $D$ from the CLF laser, the pixel or camera will observe photons from a laser path length $dx$ in the atmosphere 
and the following relation holds:
\begin{eqnarray}
dx      &=& \frac{D^2+h^2}{D/\tan(\mathrm{FOV}) - h} \qquad . \label{eq:dx}
\end{eqnarray}

\begin{figure}[h!]
\centering
\includegraphics[width=0.49\textwidth]{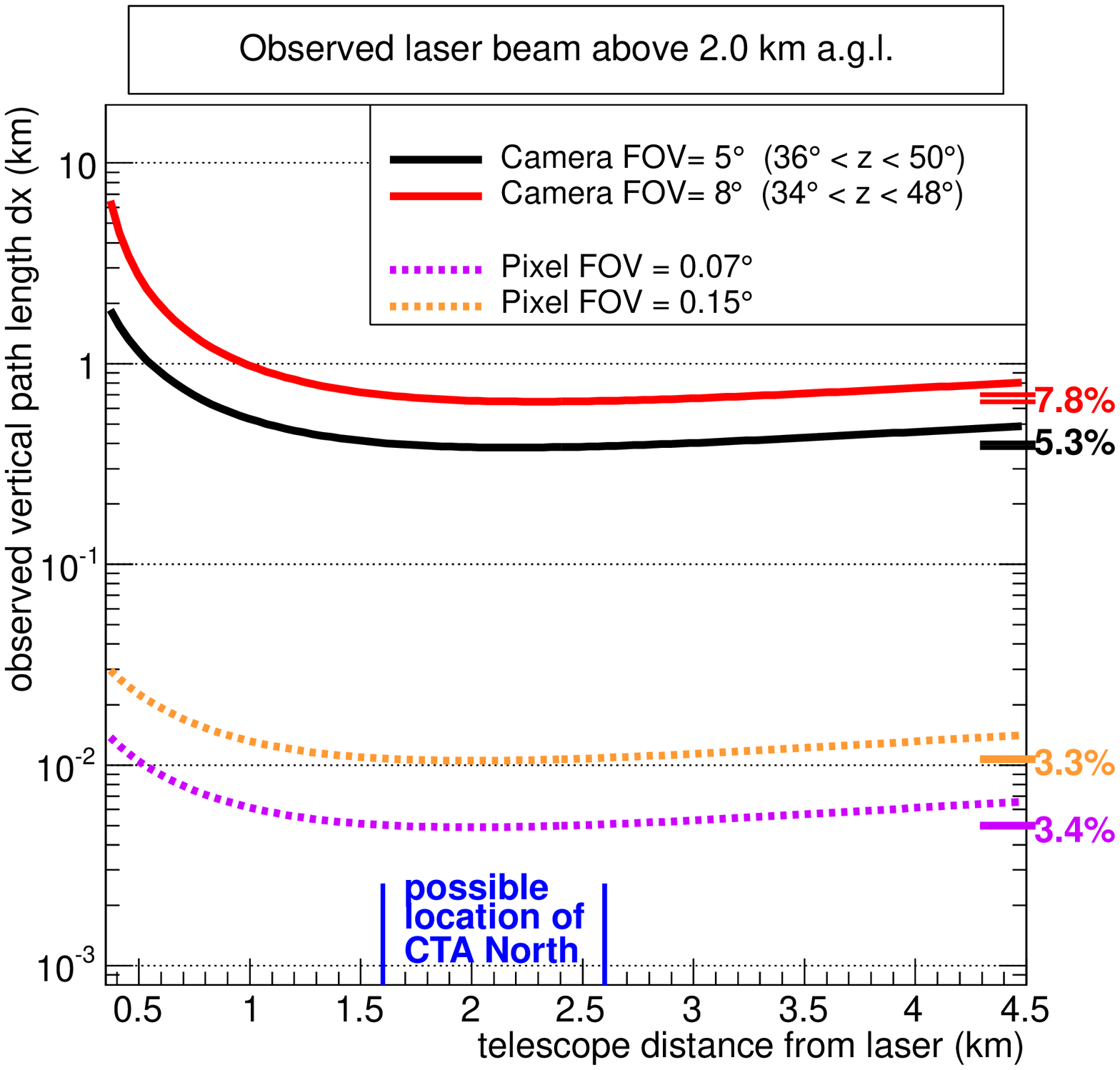}
\includegraphics[width=0.49\textwidth]{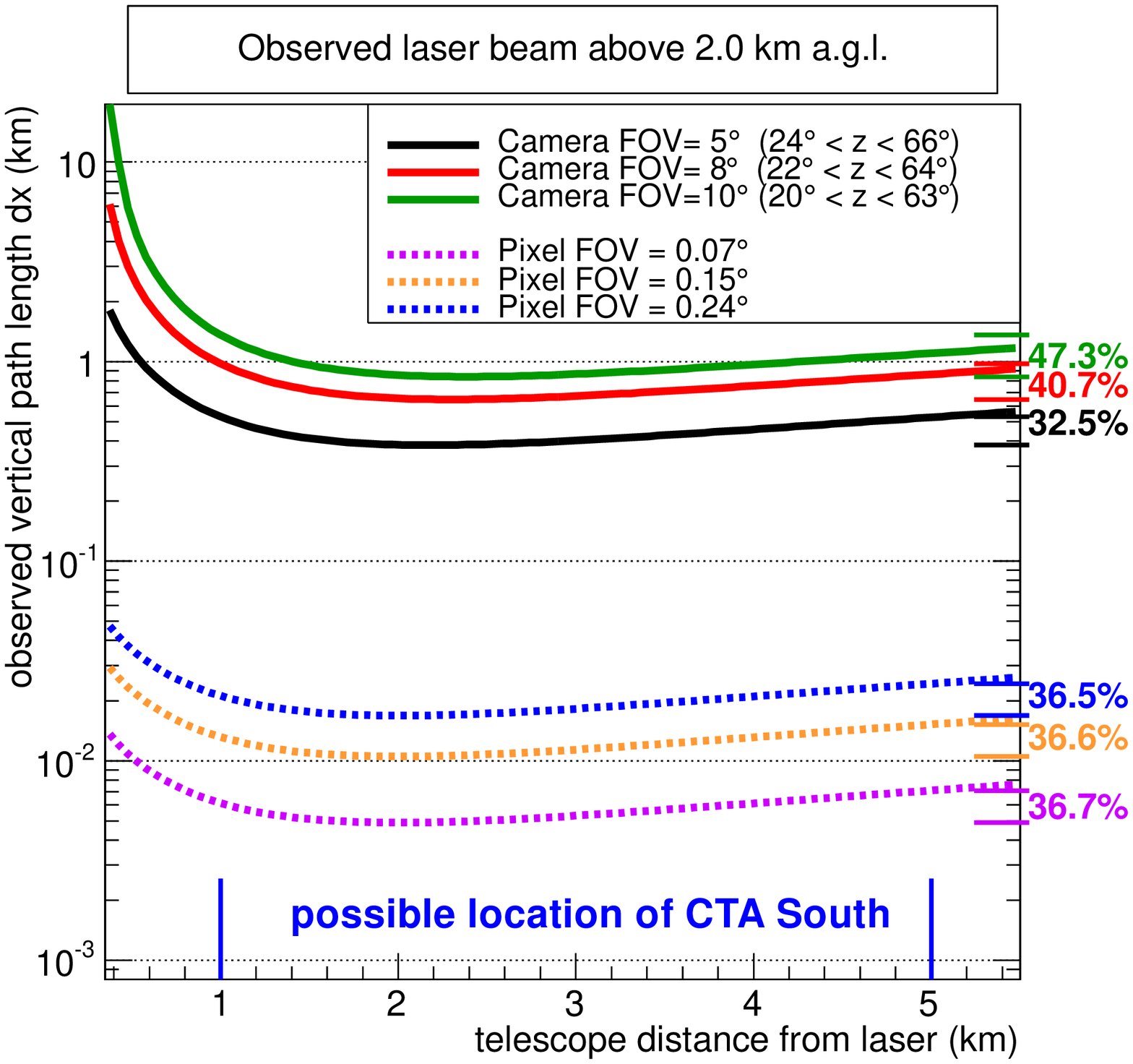}
\caption{Equation~\ref{eq:dx} plotted for different FOV values for the camera (full lines) and single pixels (dotted lines). 
At the bottom in the center, the possible locations of the 
CTA (top: Northern array, bottom: Southern array) are suggested.  
At the right side, the relative differences in path length
are printed, between the minimum and maximum value of \textit{dx} within the suggested position of the CTA.
The legend shows also the corresponding range of zenith angles ($z$). \label{fig2}}
\end{figure}

\noindent
Figure~\ref{fig2} shows the behavior of \textit{dx} for different discussed FOV design values~\cite{cta}, 
when plotted against the distance $D$.
There is a broad minimum of \textit{dx} found between around 1.6 and 2.6~km, which could be a possible location of the CTA-North with respect to the laser 
(Figure~\ref{fig2} top). 
The observed path lengths of the laser beam differ by less than 8\,\%, for all telescopes of a same type, even in the case of an 8$^\circ$~wide-field camera. 
The suggested solution for the position of the CTA means that the telescope closest to the laser facility observes the laser track under a zenith angle of around $33^{\circ}$, 
depending slightly on the camera FOV, and the farthest telescope points to the laser beam under a zenith angle of around $50^{\circ}$. 
For the more extended Southern array, telescope positions from 1--5~km for the CLF are proposed, yielding observed laser beam length differences between 
30\% and 50\%, the latter for the case of the small telescopes with a 10$^\circ$ FOV (figure~\ref{fig2} bottom).



The resulting typical transit times of the light pulses for the currently used design FOV values for the 
different telescope types of the CTA range then from tens to hundred nano-seconds for the individual pixels 
and several micro-seconds for the entire camera. These numbers can only be reduced by lowering the height of the observed laser path, at the 
cost of a bigger contribution of aerosols to the scattering of light into the camera. 


\section{Achievable precision}

The statistical precision of this calibration procedure depends mainly on the number of photo-electrons accumulated in those pixels which are used 
for the signal extraction. 
Rayleigh scattering calculations of the laser light show that a factor of at least 20 between signals from different telescopes must be assumed. 
In order to ensure that a Gaussian approximation of the photo-electron statistics is precise enough, 
the laser output power should be adjusted such that a pixel of the telescope which receives the faintest light pulse, gets at least 30--50 photo-electrons. 
On the other side of the array, a camera pixel will then receive on average at least 600-1000 photo-electrons.
If 10 rows of pixels can be used until the signal ranges 
out of the recorded memory depth, statistical uncertainties of 5--6\% for the furthest telescope and around 1\% for the closest telescope 
can be achieved for one laser shot. About a hundred shots are needed to reduce the statistical error to below 1\% for all telescopes.
Additionally, the individual pixel calibration factors
will have an uncertainty of a maximum of 5\% each, adding together another 1.5\% uncertainty. We will assume 2\% statistical uncertainty 
for here on.

Apart from that, the achievable systematic precision of the CLF calibration procedure is limited by 
the FOV overlap,
the stability of the beam direction,
the precision with which atmospheric temperature and pressure are known at the scattering point, 
the limited knowledge of the vertical optical depth, 
the aerosol volume scattering cross section, 
the bandwidth of the signal amplification chain, 
leakage of background light into the recorded laser path, 
residual polarization of the laser beam, 
and possible spectral contamination of the laser wavelength by other harmonics, if a Nd:YAG laser is used. 
Finally, the laser output power is known to a limited precision. 
Commercially available standard lasers show an energy stability of $<2\%$, commercial Joule meters about 4\%. 
This value can be improved by the use of more than one independent (pyroelectric and photo-diode) probes to values of $<2\%$ \cite{auger2011}.

\begin{table}[htp]
\begin{scriptsize}
\begin{tabular}{l|l|l}
Source uncertainty          & size      & type of calibration affected   \\   \hline
Statistical                 & 1--2\%    &  ABS, I-TEL        \rule{0mm}{3mm}  \\
Laser power                 & 2\%       & ABS, TIME         \rule{0mm}{3mm}\\
FOV overlap correction      &  $<$10\%  & ABS, I-TEL, I-TYPE   \rule{0mm}{3mm} \\
(depends on telescope focus) &          &       \\
Beam direction    & 1\%        & ABS, TIME, I-TEL, I-TYPE       \rule{0mm}{3mm}\\
Atm. density at scattering point  & 1\%   & ABS, TIME  \rule{0mm}{3mm}\\
(depends on atmospheric  &        &  \\
 monitoring equipment)   &       &   \\
Vertical optical depth   & 1--2\%  & ABS, TIME, I-TEL   \rule{0mm}{3mm}\\
(clearest nights,      &         &    \\
 Lidar required )      &         &    \\
Aerosol scattering cross-section   & 1--2\% & ABS, TIME, I-TEL \rule{0mm}{3mm}\\
(depends on atmospheric  &       &    \\
 monitoring equipment )  &       &    \\
Limited bandwidth correction  &  0--10\%  & ABS, I-TYPE   \rule{0mm}{3mm}\\
(depends on              &          &     \\
electronic coupling)     &          &     \\
Background light        & 2--8\% & ABS, TIME, I-TEL, I-TYPE \rule{0mm}{3mm}\\
(depends on FOV: lower value  &        &   \\
for carefully selected pointings) &        &   \\
Beam polarization  & $<$2\% & ABS, I-TEL   \rule{0mm}{3mm}\\
Spectral contam. & $<$1\% & ABS     \rule{0mm}{3mm}\\
(only for Nd:YAG lasers )   &      &      \\
\hline
\end{tabular}
\end{scriptsize}
\caption{List of sources of systematic errors for the CLF calibration types: absolute calibration of the whole array (ABS), 
time evolution of the telescope response for each individual telescope (TIME), inter-telescope calibration (\mbox{I-TEL}) and 
calibration of the three different telescope types w.r.t. each other (I-TYPE).  \label{tab:systerr}}
\end{table}

Table~\ref{tab:systerr} shows the assumed magnitude of these systematic uncertainties and which type of calibration they affect. 
All numbers reflect best guess estimates which will probably be reduced with time and experience and additional hardware 
to measure each contribution separately. Including statistical uncertainties, an absolute calibration of about 4--5\% uncertainty 
seems possible, if the camera hardware is adapted to the requirements of a CLF. Additionally, each monitoring point will fluctuate by the same amount. 
Note especially that the numbers given for the inter-telescope 
and inter-telescope-type calibration reflect the uncertainty for each calibration run and can probably be reduced by applying 
the calibration procedure throughout different nights. 

\section{Hardware considerations}

A CLF will do its work only if 
several hardware requirements are met on the side of the CTA telescopes:

Each telescope should have the possibility to focus to a distance of about 3~km or lower.
Since each pixel along the viewed laser path receives the signal at different times, different readout depths 
for each channel along the signal axis are required. 
The front-end electronics needs to be able to amplify and electronically transmit longer signal pulses
in the same way as the short Cherenkov light pulses. 
A signal digitization mode must be implemented which allows to integrate the entire pulse from CLF runs.
Finally, an external trigger needs to be installed for each telescope individually to trigger the CLF readout such that the central pixels'  
signals get correctly centered in time or, alternatively, a differential trigger must be able to trigger the readout of each pixel (or cluster of pixels) at the 
correct time delay w.r.t. to the previously hit pixel (or cluster).

\begin{table}[htp]
\centering
\begin{tabular}{|c|c|c|}
\hline
    & \multicolumn{2}{|c|}{Minimum pulse width registration capability} \\
\hline
    & Extension 1~km & Extension 4~km \\
\hline
LST &   15~ns        &  25~ns  \\
MST &   30~ns        &  50~ns  \\
SST &   50~ns        &  80~ns  \\
\hline
    & \multicolumn{2}{|c|}{Mean time delay between clusters} \\
\hline
    & Extension 1~km & Extension 4~km \\
\hline
LST &   42~ns        &  70~ns  \\
MST &   84~ns        & 140~ns  \\
SST &   140~ns       & 220~ns  \\
\hline
\end{tabular}
\caption{Minimum requirements
for the current design parameters of the 
large-size telescope (LST), the medium-size telescope (MST) and the small-size telescope (SST).
The mean time delay refers to 7-pixel clusters in a hexagonal camera geometry. \label{tab:minimum} }
\end{table}

Based on these points, minimum requirements for the telescopes cameras can be set.
Once the optics are fixed and the final extension of the array known, a minimum pulse width recording capability can be derived for 
an optimal case ($h = 2$~km) and an absolute minimum case ($h = 1$~km). The results are shown in tables~\ref{tab:minimum} and~\ref{tab:optimum}, respectively.
One can see immediately that the obtained values are most critical for the small-size telescope, 
due to the larger FOV and to the physical extent of the Southern array. 
Using several CLFs, located at each side of the array, could help to reduce the 4~km extension requirements to those obtained for a 2~km extension array. 
This option would also reduce systematic errors related to the atmospheric transmission.



This allows derivation of an absolute minimum requirement for the correctly amplified and registered pulse width of: 15, 35 and 60~ns for 
the Southern array of large-size, medium-size and small-size telescopes, respectively, and 15 and 30~ns for the Northern array. 
Concerning the triggerable time delay between 
two pixel clusters, the absolute minimum requirements are: 40, 100 and 170~ns for the Southern array and 40, 80 and 140~ns for the Northern array for 
the large-size, medium-size and small-size telescopes, respectively.

\begin{table}[htp]
\centering
\begin{tabular}{|c|c|c|}
\hline
    & \multicolumn{2}{|c|}{Minimum pulse width recording capability} \\
\hline
    & Extension 1~km & Extension 4~km \\
\hline
LST &   27~ns        &  33~ns  \\
MST &   60~ns        &  73~ns  \\
SST &   96~ns        & 120~ns  \\
\hline
    & \multicolumn{2}{|c|}{Mean time delay between clusters} \\
\hline
    & Extension 1~km & Extension 4~km \\
\hline
LST &   75~ns        &  90~ns  \\
MST &  170~ns        & 200~ns  \\
SST &  270~ns        & 340~ns  \\
\hline
\end{tabular}
\caption{Requirements for an optimum case
for the current design parameters of the 
large-size telescope (LST), the medium-size telescope (MST) and the small-size telescope (SST). 
The mean time delay refers to 7-pixel clusters in a hexagonal camera geometry. \label{tab:optimum} }
\end{table}

\section{Discussion and conclusions}

How does the precision and the impact of a CLF-based calibration scheme compare with other techniques?
We do not consider calibration methods for individual telescopes with systematic uncertainties in the range of 10--15\% 
or techniques which demand considerable down-time of the array to calibrate it. 





Elaborate methods comparing images recorded from local muons with those from simulations 
have been developed~\cite{rose,puhlhofer,leroy,goebel,bolz}, reaching a precision down to a few percent. 
To achieve this, shadowing of the mirrors by the camera or masts needs to be simulated thoroughly, 
and the effect of the different emission spectra 
(starting from 250~nm wavelength in the case of muons, while gamma-ray shower light is typically absorbed below 300~nm) needs to be corrected for~\cite{humensky}.
Muons lose only 3.5--4\% of the light in the range from 300 to 600~nm from the point of emission to the photo detector in the camera~\cite{bolz} and are hence 
much less affected by the corresponding atmospheric uncertainties, if compared to a CLF. It has to be clear however, that calibration methods 
using muons also require some hardware adaptation from the individual telescopes trigger. Otherwise, muons may be completely lost since the may be rejected by the 
multi-telescope trigger or a too high threshold. Furthermore, it is still not clear whether the small-size telescopes will be sufficiently sensitive to register useful 
muon images for calibration. Finally, a CLF can provide calibration at distinct wavelengths, while the Cherenkov light spectrum from muons cannot be changed.  


In the low-energy range, cross-calibration with sources observed by the (much more precisely calibrated) 
FERMI satellite can yield a precision of ${}^{+5\%}_{-10\%}$~\cite{meyer,bastieri}. 
This will calibrate the energy reconstruction of the large-size and combinations of medium-size telescopes, while the small-size telescopes 
need to rely on other known spectral features at higher energies, such as the cut-off in the cosmic-ray electron spectrum, measured by PAMELA and AMS. 

Using cosmic ray images, and particularly the distributions of their image sizes and reconstructed shower impact points, 
\cite{hofmann} claims a precision of 1--2\% for inter-telescope calibration. As in the case of muons, the method seems superior in terms of precision 
and does not need any hardware adaptation, however information about the spectral sensitivity cannot be derived, and overall degradations of the array 
cannot be detected, especially if the atmosphere is not understood to the same level of precision. 

In summary, it seems useful to operate a CLF for combined and fast calibration of the array. 
Although other methods may result in a partially more precise 
calibration, these cannot provide inter-telescope, inter-telescope-type, absolute and spectral calibration at the same time. 
Since at the aimed 
performance of the CTA different calibration methods must be cross-checked with others, a CLF presents itself as a reasonable candidate.

\vspace{0.5cm}
{\bf Acknowledgements:\xspace} The authors would like to thank L. Valore for valuable comments and
discussions. 
We gratefully acknowledge support from the agencies and organizations  
  listed in this page: \\
\url{http://www.cta-observatory.org/?q=node/22}

\end{document}